# SUPERNOVAE PHOTOMETRY AT OAUNI[1]

M. Espinoza[2] and A. Pereyra[3,2]

*Draft version: June 4, 2024*

### RESUMEN

Analizamos los datos fotométricos de nueve supernovas en los filtros *V*, *R* e *I* que se obtuvieron durante las campañas observacionales de OAUNI en 2016, 2017 y 2023. Para investigar la evolución posterior a su punto máximo de brillo, se compararon las magnitudes calibradas de las supernovas observadas con sus respectivas curvas de luz disponibles en la literatura. En algunos casos, se usó el diagrama de diagnóstico color-color de supernovas para determinar nuestras fechas de observación y ubicarlas correctamente en las curvas de luz. Para este prop ?osito también fueron de ayuda la utilización de plantillas de curvas de luz de supernovas, as´ı como supernovas de referencia. Este trabajo permitió verificar la factibilidad de realizar fotometría astronómica de precisión en el OAUNI.

### ABSTRACT

We analyse photometric data of nine supernovae (SNe) in filters V, R and I obtained during observational campaigns at the OAUNI site in 2016, 2017 and 2023. The calibrated magnitudes of the observed SNe were compared with their respective light curves available in the literature to study their evolution after their maximum brightness. In some cases, the supernova color-color diagnostic diagram was used to determine our observation date and correctly locate our magnitudes on the light curves. For this purpose, the use of supernova light curve templates, as well as reference supernovae, was also helpful. This work allowed us to verify the feasibility of performing precision astronomical photometry at the OAUNI.

*Key Words:* (Stars:)Supernovae: General — Techniques: Photometric

## 1. INTRODUCTION

The Astronomical Observatory of the National University of Engineering (OAUNI in spanish) begins their operations in 2015 (Pereyra et al. 2015). This facility is situated in Huancayo, 3300 meters above sea level, in the heart of the Peruvian Andes. One of the main scientific programs proposed was the supernovae photometric follow-up with several detections since then. This work presents the main SNe events observed at OAUNI site on the last years since 2016. Special care was taken into the photometric calibration process in order to contribute with useful data for the supernovae light curves of the analyzed events. Previous efforts in SNe observations in Perú include the detection of the famous SN 1987 by M. Ishitsuka and H Trigoso (private communication) at the same site of these observations and SN 2003gt (Carlos Reyes et al. 2013) observed at the southern peruvian Andes.

---

[1] Observations obtained at the Astronomical Observatory of the National University of Engineering (OAUNI) in Huancayo, Perú.

[2] National University of Engineering, Lima, Perú.

[3] Geophysical Institute of Peru, Astronomy Area, Lima, Peru.

In the following, we describe the observed SNe (Sect. 2), and the reduction process (Sect.3) including the different methods used for the calibration data. The analysis and comparison of OAUNI data with template light curves, diagnostic color-color diagrams, and data available in the literature for each event is shown in Sect. 4. Finally, our conclusions are drawn in Sect. 5.

## 2. DETECTED OAUNI SUPERNOVAE EVENTS

The OAUNI SN sample is indicated in Table 1 and Figure 1. A total of nine events were detected including four SNe Type Ia, four Type II, and one Type Ib. All the analyzed SNe are at a redshift lower than 0.04 (see Table 1). Below is relevant information about each event.

### 2.1. AT 2015dd

On December 15, 2015, AT 2015dd was found in the center of the galaxy NGC 5483 ($z$ = 0.006) by the MASTER-SAAO[4] (Gress et al. 2015). Using the SOAR telescope, three days later, it was identified as a Type Ib SN, and 2015-12-08 was determined to be the day of maximum brightness (Foley, Hounsell & Miller 2015).

### 2.2. SN 2016cvk

On June 12, 2016, the BOSS group[5] discovered SN 2016cvk. The SN was situated east of the host galaxy ESO 344-G 021 ($z$=0.010783, Parker 2016), and had a very low brightness of ∼17 mag in *V* filter. With behavior very similar to SN 2009ip, the *ASASSN* group[6] classified this SN as Type IIn. The PESSTO group (Parker 2016) further confirmed this classification.

### 2.3. AT 2016eqb

On August 1, 2016, the *ASASSN* group (Brimacombe et al. 2016) identified AT 2016eqb in the host galaxy 2MASX J23154564-0120135 ($z$=0.025308). KOSMOS[7] classified it as a Type Ia SN, with the day of maximum brightness being 2016-08-07 (Pan et al. 2016).

---

[4]Mobile Astronomical System of Telescope-Robots at the South African Astronomical Observatory, a self-detection system
[5]Backyard Observatory Supernova Search
[6]All-Sky Automated Survey for SNe
[7]The Cosmic Evolution Survey (COSMOS)

### 2.4. SN 2017erp

K. Itagaki discovered SN 2017erp on June 13, 2017 (Itagaki 2017). SALT[8] (Jha et al. 2017) classified it as an extremely young Type Ia SN, and it is located in the arm of NGC 5861 (z = 0.006174±0.000003, Theureau et al. 2020). This SN was the special interest by the relationship between its non-homogeneous composition and its light curve, as well as the peculiar reddening of its spectral lines in the near-ultraviolet range.

### 2.5. AT 2017erv and AT 2017eve

The *ASASSN* group found AT 2017erv on 2017-06-13 in AM 1904-844 (z=0.017035). A few days later, on June 19, 2017, AT 2017eve was also found on the same images in GALEXASC J184352.21-562927.7 (z=0.031, Nicholls, Brimacombe & Cacella 2017). On June 20, 2017, (Uddin et al. 2017) categorized both SNe as Type Ia SNe, with a phase from maximum brightness of -2 days for AT 2017erv and +2 days for AT 2017eve.

### 2.6. SN 2023gfo

On 2023-04-19, SN 2023gfo was detected in NGC 4995 (z=0.0058) by the ATLAS[9] system. Additionally, the same field was observed four days prior to this detection, but no sign of the SN was found. According to Moore et al. (2023), this event would suggest that the SN was in its growth phase. Lick Observatory classified it as an SN Type IIP with a spectrum remarkably similar to SN 1999gi (Fulton et al. 2023).

### 2.7. SN 2023ijd

The *ASASSN* group found SN 2023ijd in NGC 4568 (z = 0.007446, Stanek 2023) on 2023-05-14. It was classified as a Type II SN (Perley 2023).

### 2.8. SN 2023ixf

SN 2023ixf was classified as a Type II SN in its early stages of life (Perley & Gal-Yam 2023) after K. Itagaki discovered it on May 19, 2023, in M101 (z = 0.000804, Itagaki 2023). Over the past few decades, SN 2023ixf has been considered the closest Type II SN. Subsequent reports of earlier sightings, following the discovery, helped narrow down the explosion date (Smith et al. 2023; Filippenko, Zheng & Yang 2003) to a 20-hour window between May 18 and 19. SN 2023ixf was later reclassified as Type II-L (Bianciardi et al. 2023).

---

[8]The Southern African Large Telescope
[9]Asteroid Terrestrial-Impact Last Alert System

## 3. OBSERVATIONS, REDUCTIONS AND CALIBRATIONS

All observations mentioned here were collected using the OAUNI telescope (Pereyra et al. 2015) during the 2016, 2017, and 2023 observation campaigns. These observational runs typically take place in the months of May through September. In only one instance (AT 2015dd), a single observation was made in January. The OAUNI telescope has a Cassegrain type optical tube with Ritchey-Chrétien design and a primary mirror with a diameter of 0.51 m and f/8.2. A front-illuminated CCD STXL-6303E with 3072×2048 pixels$^2$ and 9$\mu$m/pixel served as the detector. A field-of-view of $\sim$ 23$'$ × 15$'$ and a plate scale of 0.45$''$/pixel are produced by this detector and the optical system's focal ratio. For the scientific objects, multicolor photometry was made possible via a *UBVRI* filter wheel. The record of observations made during the campaigns is displayed in Table 2. Column (1) shows the name of the SN, column (2) presents the local observation date, column (3) indicates the filters used, column (4) displays the number of images obtained in each filter, and column (5) shows the total integration time for each case. Column (6) presents the mean airmass during each sequence. In total, data from nine SNe are presented, with three different SNe observed each year. The individual integration time for one measurement is 20 seconds, and the total time for stacking images ($N$×20s) ranges from 600 to 1400 seconds.

Using standard corrections for dark current and flatfield, we used IRAF[10] for image reduction. Aperture photometry was extensively used with typical instrumental magnitude error of tens of milimagnitude for the magnitude range of our sample (typically, between 11.3 to 17.1 mag). The first step in the calibration process was to find stars in each stellar field that matched both our images and the UCAC4 photometric catalog (Zacharias et al. 2013). These stars, listed in Table 3, were then used as comparison stars for every SN analyzed. We utilized two methods to determine the corrected value of the SN brightness using this data. The first method ($m_1$), involving equations 1, 2, and 3, was used to represent the transformation of the instrumental magnitudes ($v$, $i$, and $r$) to the calibrated magnitudes ($V$, $I$, and $R$) by obtaining a single zero point ($v_0$, $i_0$, and $r_0$). This method is useful when only one filter is available for measurements and was used for all the objects in our sample.

$$V = v_0 + v \tag{1}$$

$$R = r_0 + r \tag{2}$$

$$I = i_0 + i \tag{3}$$

Using the transformation equations, the second method ($m_2$) involves the zero points ($v_0$, $r_0$, and $i_0$), the linear dependence ($v_1$, $r_1$, and $i_1$), and the coefficients ($v_2$, $r_2$, and $i_2$) for the color terms ($v - r$ or $v - i$) of the objects under consideration. The equations 4, 5, and 6 illustrate these transformations. This calibration method is more robust but is necessary multicolor photometry for its applicability. It was used on five of the sample's objects.

---

[10]Image Reduction and Analysis Facility hosted by the National Optical Astronomy Observatories in Tucson, Arizona

$$V = v_0 + v_1 \times v + v_2 \times (v - r) \tag{4}$$

$$R = r_0 + r_1 \times r + r_2 \times (v - r) \tag{5}$$

$$I = i_0 + i_1 \times i + i_2 \times (v - i) \tag{6}$$

Our findings for the calibrations of the SNe magnitude of our sample, $m_1$ (column 7) and $m_2$ (column 8), are displayed in Table 2.

## 4. ANALYSES

Figure 2 displays $m_1$ and $m_2$ for any scenario in which both calibrations are provided for the same object so that the two calibration techniques can be compared. With the exception of 2016cvk, all examples have high point-to-point correlation, and the typical residual between $m_1$ and $m_2$ is 0.096 ± 0.064 mag. When two calibrations are available, we will utilize $m_2$ for the analyses in the following; in other circumstances, we will use $m_1$.

### 4.1. AT 2015dd

When a SN reaches its maximum brightness, color-color diagnostic diagrams can be used to confirm how old it is (Poznanski et al. 2002). Based on the Type Ib SN classification of AT 2015dd and the near-zero redshift of the host galaxy (see to Table 1), Figure 3 illustrates the temporal behavior of a well-studied Type Ib SN at various stages of its life for $z = 0$. Using the $m_2$ calibration (Table 2), we computed the colors $V - R$ and $R - I$ for AT 2015dd. The values $V - R$ = 0.864 ± 0.047 and $R - I$ = 0.252 ± 0.035 are also displayed in Figure 3. The location of AT 2015dd on the diagram indicates a period of 20-40 days after the maximum of brightness, although it is not sufficient to explicitly validate the age of the supernova. Nevertheless, our observations are ∼32 days after the maximum, taking into account the day of the explosion on 2015-12-08 (Foley, Hounsell & Miller 2015).

### 4.2. 2016cvk

Based on the information found in Table 2, we calculated the $m_2$ colors for SN 2016cvk, which are $V - R$=0.045 ± 0.004 and $R - I$=0.002 ± 0.005 on the Type IIn SN's z=0 diagnostic diagram (refer to Figure 4). This SN 2016cvk is located almost exactly on the line that the diagram's days 3 through 13 encompass. It implies that when OAUNI observed this SN, it was still very young. On the other hand, we used a template (Nugent, Kim & Perlmutter 2002), literature data (see Table 4), and OAUNI $m_2$ data (see Table 2) to generate the light curve in $V$ filter for this SN. To determine whether the brightness of this SN behaves similarly to the average brightness of SNe of the same type, the Nugent's template is shown. In order to compare the data, we must fit all of the data to the same reference system because Nugent's template plots the peak of brightness in the $B$ filter of the time coordinate. First, we use equation 7 to convert the numbers at the template's peak to a polynomial.

$$m(t) = \sum_{i=0}^{5} m_i \times (t - t_0)^i \qquad (7)$$

Using data from the SN close to the peak, we modify $m_0$ and $t_0$ in this polynomial to determine the values that present the minor residual. We divide the duration of the data by 1+z to subtract the $z$ contribution of the host galaxy (see Table 1) before substituting this data. In order to modify the template to fit the displayed data, we acquired the final vector ($t_0$; $m_0$) after changing the SN data (see Table 5). Since the maximum brightness date is unknown, we set the beginning $t_0$ value for the fitting using the information provided by the diagnostic diagram (Figure 4). The ultimate outcome of the development of the light curve, where the data follows the template, is shown in Figure 5. The diagnostic diagram places this event around the day 10 after the peak, however the OAUNI data is situated around the day ∼-12 before the peak.

### 4.3. *AT 2016eqb*

Since the maximum brightness date for AT 2016eqb is known (Pan et al. 2016), we have plotted it alongside the OAUNI $m_2$ data (see Table 2) and the available literature data (see Table 4) using the minor residue method to determine $m_0$ only and fit the type Ia SN Nugent's template in $V$ filter. Since the host galaxy's $z$ value is known in this instance as well (see Table 1), we used all of the data points that are near the peak to calculate the value of $m_0$. Nugent's template in $V$ filter with the appropriate $t_0$ and $m_0$ adjustments, as well as the OAUNI photometry of AT 2016eqb, are displayed in Figure 6 (see Table 5).

### 4.4. *2017erp*

We utilized values close to the peak of the light curves (15 days before and after the maximum) in both filters ($V$ and $R$) to fit the type Ia SN Nugent's template because of the large amount of available literature data. The minor residue approach was used to compute $m_0$, similar to the last supernova, since the date of the maximum brightness is known ($t_0$ = 2457934.9 JD, Brown et al.

2016). We got distinct values of $m_0$ (see Table 5) for each filter, accounting for $z$ from its host galaxy (see Table 1). Using the OAUNI $m_2$ data (see Table 2) and the available literature data for SN 2017erp (Brown et al. 2016), Figure 7 and 8 display the light curves in *V* and *R* filters with their corresponding templates. Both charts demonstrate how well the OAUNI data match the data from UVOT, LCO, and AZT[11]. Even while the *V* filter's light curve up to day ∼30 follows the template's trend, as the days pass, the discrepancy gets larger. The *R* filter's light curve, on the other hand, traces the trend both before and after the maximum brightness.

### *4.5. AT 2017erv*

The AT 2017erv light curve in *V* and *R* filters, together with the type Ia SN Nugent's templates, is displayed in Figure 9 using OAUNI $m_1$ data (see Table 2) and the available literature data (see Table 4). In order to fit template in filter *R*, we used the host galaxy redshift information (see Table 1) and the date of the maximum brightness (Uddin et al. 2017). Since OAUNI data is in R filter, the template in this filter was used as a reference to plot the template in V filter. This is a feature of Nugent's template, as each template was created using a correlation filter-to-filter. To do this, we found $m_0$ with the minor residue by substituting OAUNI $m_1$ data in the polynomial fit for the template in filter *R* (see Table 5). Next, we used Nugent's SN Ia template light curve to compare the OAUNI findings in the *R* filter (Table 2) with the values found by *GAIA* and *ASASSN* in the *V* and *G* filters, since this SN was categorized as a Type Ia (Uddin et al. 2017). The outcome of fitting Nugent's template to OAUNI *R* filter data is displayed in Figure 9. Furthermore, we can see that, in contrast to the template in the *V* filter, the *ASASSN* data exhibits a continuous rise in brightness before the peak, whereas the *GAIA* data has a faster fall rate.

### *4.6. AT 2017eve*

We have used Nugent's templates in the *R* and *V* filters, same like in the previous SN. We employed OAUNI $m_1$ data (see Table 2) in the minor residue technique to obtain the value $m_0$ solely (see Table 5), taking into account its categorization as a Type Ia SN, the date of the peak, and the host galaxy $z$ (see Table 1). This fitting enables us to see the *ASASSN* data dispersion around the maximum brightness date in Figure 10. However, *GAIA* data show a different behavior from the preceding SN, with a smaller fall rate than the template.

---

[11]Shamakhi Astrophysical Observatory.

*4.7. SN 2023gfo*

The *V-R* and *R-I* colors for the single OAUNI multicolor photometry data (see to Table 2) of SN 2023gfo have been computed and are shown in Figure 11. Although its location in the diagnostic diagram is not precise enough to determine the observational period, it indicates that this SN may have occurred between days 34 and 42 following the maximum brightness. The telegraph of its discovery (Moore et al. 2023), which highlights the fact that the ATLAS system found no evidence of this event three days earlier in the same area, despite the SN being discovered on May 19, supports this view. The type IIp SN 2004et data (Sahu 2006) and the OAUNI $m_1$ data (see Table 2) of four consecutive months in filter *R* are plotted together in Figure 12. Since SN 2004et's data has already been adjusted for its peak brightness time, it can be used as a kind of template. Thus, in order to get the offset vector ($t_0$; $m_0$), we fitted the SN 2004et light curve to a polynomial. Next, we got its equivalent vector (see Table 5) from the minor residue approach after substituting the OAUNI $m_1$ data in the SN 2004et political fitting. During the initial three months, OAUNI detected a 0.6 mag decline in magnitude while SN 2004et remained within the same range. In the last month, data could set SN 2023gfo in the end of the plateau phase. The OAUNI data indicates a higher fall rate compared to prior months, which is consistent with SN 2004et. Because the first night of observation was located on day 39 using the approach of least residual, this result supports the information provided by the diagnostic diagram.

*4.8. SN 2023ijd*

We used OAUNI $m_1$ data (see Table 2) and ZTF (The Zwicky Transient Facility, Bellm et al. 2016) data[12] to plot the light curve of SN 2023ijd in Figure 13. We have shown OAUNI and ZTF data with the SN 2004et light curve in filter *R* with *z* = 0.0002 (type IIp), just as the previous SN with a similar type. We plotted the SN 2004et light curve against the SN 2023ijd data by following the same procedures as in the prior case. In this instance, we used ZTF data close to the peak to estimate the vector ($t_0$; $m_0$) with the minor residue because the date of the maximum brightness is unknown (see Table 5). The data shows a steady phase of decrease with a magnitude variation of $\pm$ 0.1 mag from +10 to +60 days following the peak. This light curve phase may represent the plateau creation phase of SN 2004et.

*4.9. SN 2023ixf*

The last SN studied was the closer and brilliant SN2023ixf. This fact is reflected in the wide available literature data up to +225 days from AAVSO[13]. As usual, we corrected the temporal values using z of its host galaxy (see

---

[12]https://lasair-ztf.lsst.ac.uk/objects/ZTF23aajrmfh/
[13]The American Association of Variable Star Observers (AAVSO)

Table 1), and we interpolated the peak date $t_0$ = 2460094.0 JD. This fact is supported by (Filippenko, Zheng & Yang 2003), where they set $t_0$ = 2460094.2 JD. The OAUNI $m_1$ data (see Table 2) for the eight nights of observation of SN 2023ixf match the trend of the light curve between days +0 and +50, as can be observed in Figure 14. Following these days, the brightness shows an increase in slope, declining in 20 days from 12 mag to 13.5 mag. At last, the SN resumes its gradual decline phase.

## 5. CONCLUSIONS

Using the equipment available at OAUNI, the reduction of images from different SNe was accomplished effectively. These nine SNe were observed in the *V*, *R*, and *I* filters over a total of 43 nights, confirming the good quality of the images. This was made possible by the fact that on several observation nights, we were able to maintain an uncertainty of less than 0.09 mag despite the presence of high air masses.

Data calibration for $m_1$ and $m_2$ method was carefully examined in each filter. As a result, the agreement between both methods has an average precision of 0.096 $\pm$ 0.064. Diagnostic diagram was used to evaluate the position fo three supernovae after the maximum brightness date. The position of the supernovae found in this diagram was corroborated with different sources that recorded the date on which these events reached their peak magnitude. Through the construction of light curves using several templates and a comparison of literature and OAUNI data, we were able to study the brightness behavior of each SN. The fitting of these templates (for SNe Types Ia and IIn) and SN2004et light curve (for Type IIp) was carried out carefully using the available data close to the peak and seeking to ensure that the residual was as small as possible (> 0.001). The viability of conducting precise astronomical photometric programs at the OAUNI site is validated by this work.

## 6. ACKNOWLEDGMENTS

The authors are grateful for the economic support from Concytec (Contrato N$^o$ PE501081907-2022-PROCIENCIA, Contrato 133-2020 Fondecyt). Special thanks to the Huancayo Observatory staff for the logistic support and J, Tello, M. Zevallos, J. Ricra, R. Santacruz, D. Alvarado and E. Torre for their collaboration with the observations.

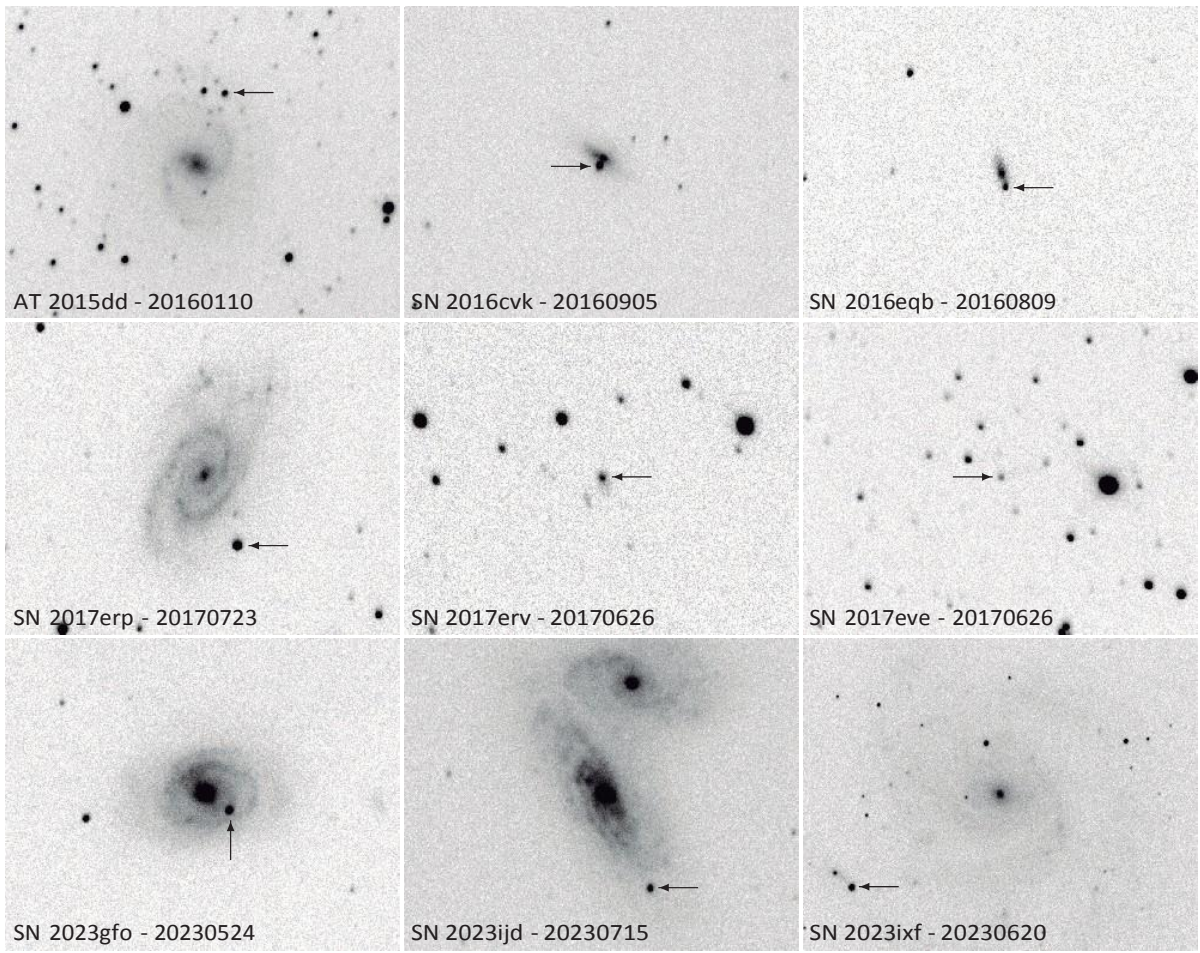

Fig. 1. OAUNI SNe smaple in *R* filter. Each frame indicates the observed SN (arrow) and the observation date. The individual FOV is $4\rlap{.}'1\times3\rlap{.}'2$, except SN 2023ixf with $9\rlap{.}'6\times7\rlap{.}'7$. North is top and East is left.

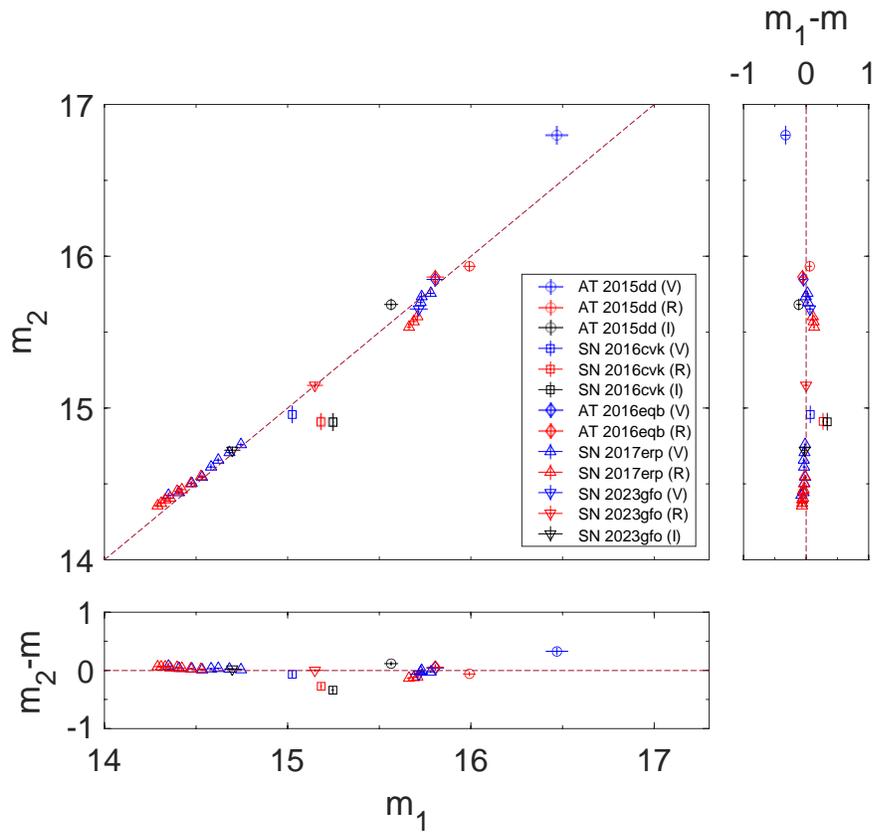

Fig. 2. Point-to-point correlation between $m_1$ and $m_2$ calibration methods for OAUNI SNe sample. Instrumental magnitudes in *V* (blue dots), *R* (red dots) and *I* filters (black dots) are shown with different symbols for each supernova. The residuals for each calibration respect the perfect positive correlation (dashed line) are indicated below and the right.

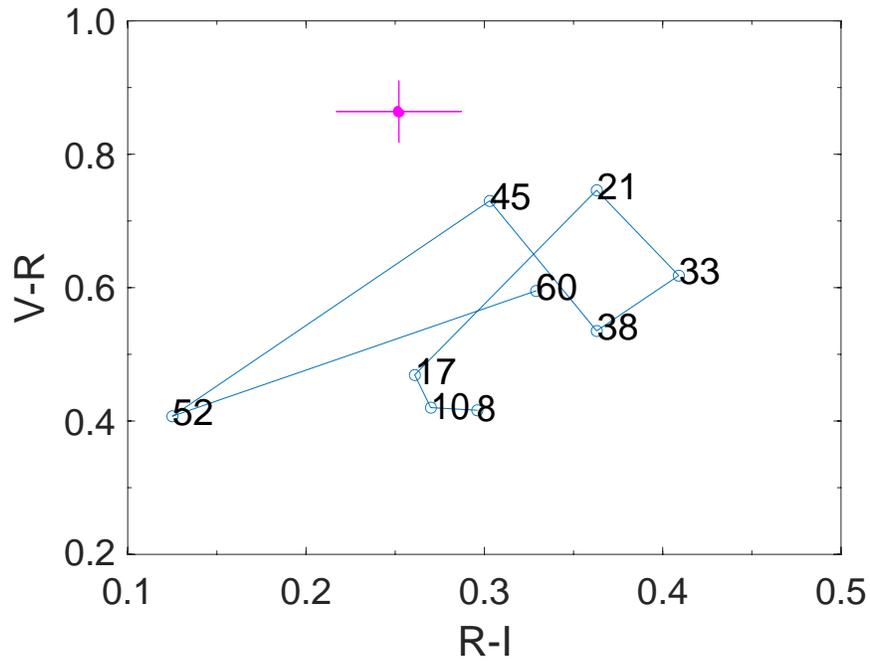

Fig. 3. $V - R$ vs. $R - I$ color diagram of Type Ib SN with $z = 0$ (blue line), adapted from Poznanski et al. (2002). Numbers indicate days after maximum of light. OAUNI colors for AT 2015dd using $m_2$ calibration are also indicated (magenta dot).

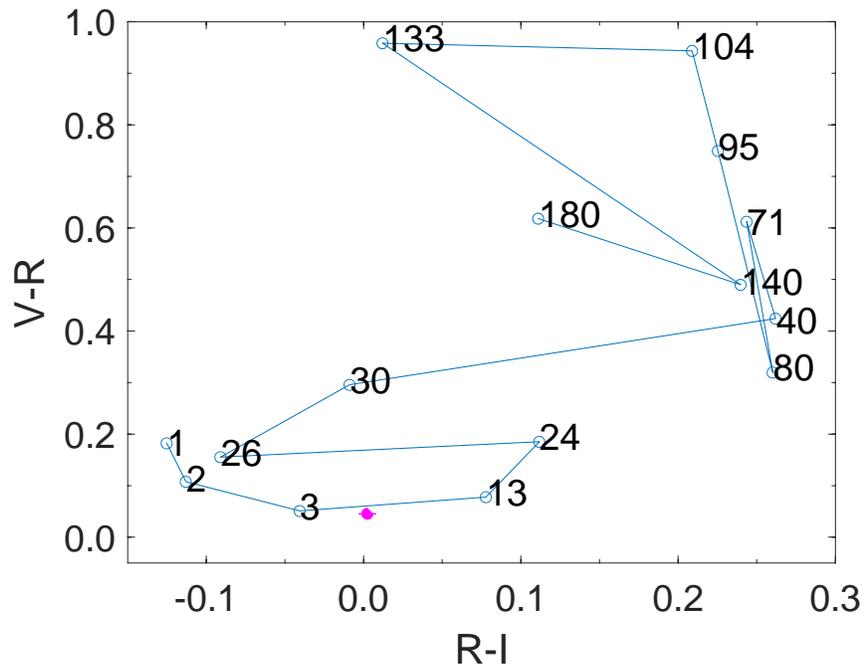

Fig. 4. $V-R$ vs. $R-I$ color diagram of Type IIn SN with $z = 0$ (blue line, adapted from Poznanski et al. 2002). Numbers indicate days after maximum of light. OAUNI colors for SN 2016cvk using $m_2$ calibration are also indicated (magenta dot).

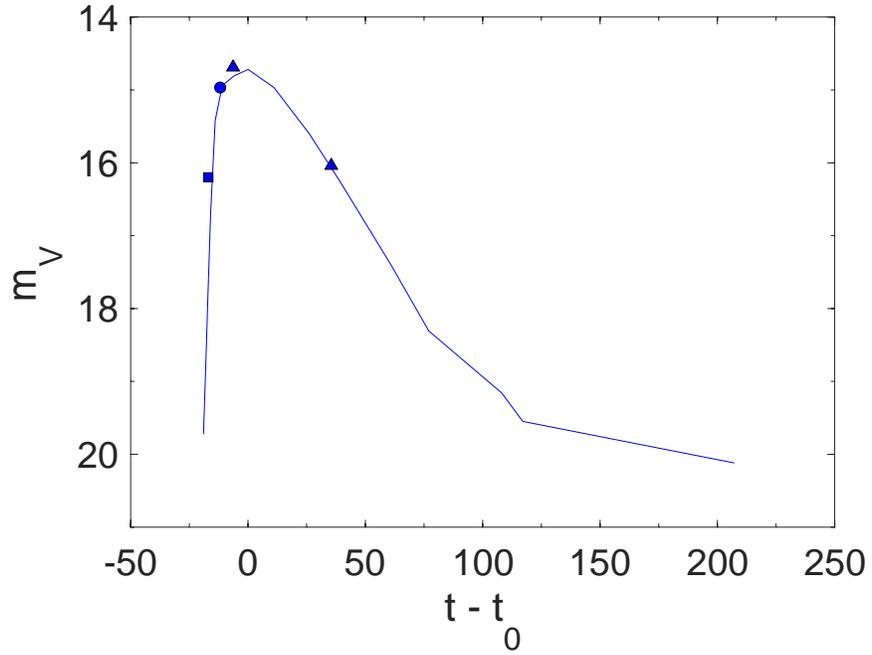

Fig. 5. SN 2016cvk light curve in *V* filter. Template light curve for Type IIn SN with $z = 0$ (blue line, Nugent, Kim & Perlmutter 2002) with *ASASSN* data (blue square, Brimacombe et al. 2016), Dr. Kato's data (blue triangle, http://ooruri.kusastro.kyoto-u.ac.jp/mailarchive/vsnet-recent-sn/6612) and OAUNI $m_2$ data (blue dot, Table 2). Time offset is $t_0 = 2457648.5$ days.

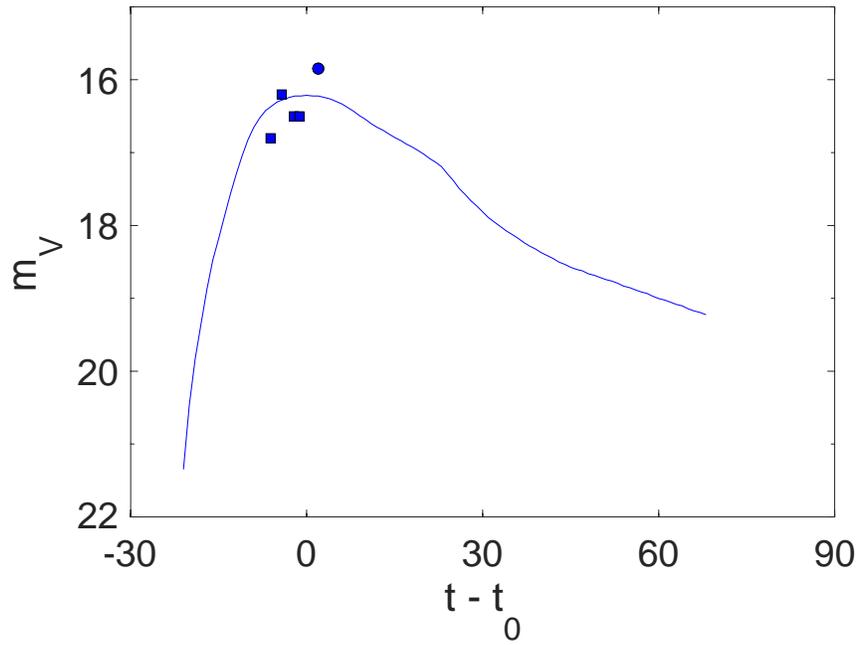

Fig. 6. AT 2016eqb light curve in *V* filter. Template light curve for Type Ia SN with *z* = 0 (blue line, Nugent, Kim & Perlmutter 2002) with *ASASSN* data (blue square, Brimacombe et al. 2016) and OAUNI $m_2$ data (blue dot, Table 2). Time offset is $t_0$ = 2457607.5 days.

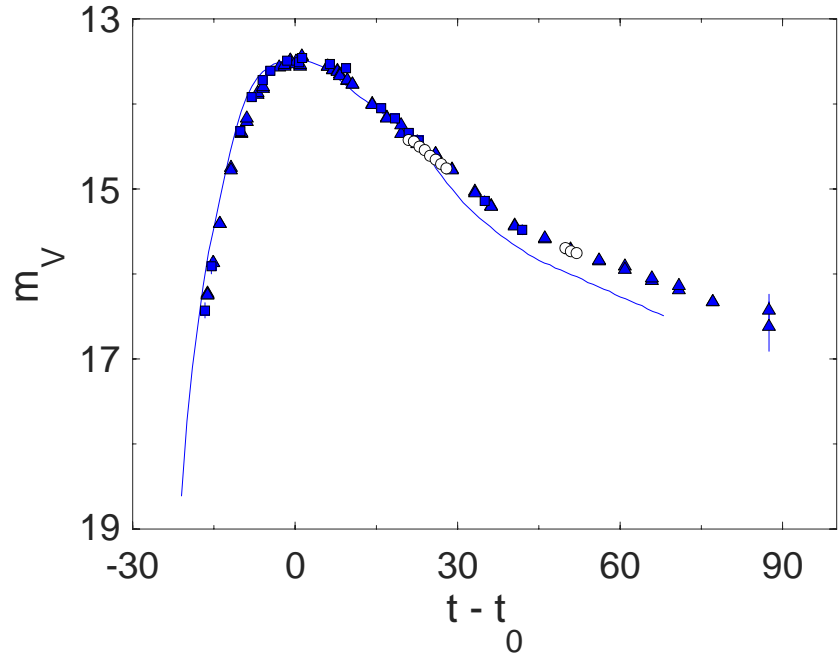

Fig. 7. SN 2017erp light curve in *V* filter. Template light curve for Type Ia SN with $z = 0$ (blue line, Nugent, Kim & Perlmutter 2002) with UVOT data (blue squares, Brown et al. 2016), LCO data (blue triangles, Brown et al. 2016) and OAUNI $m_2$ data (white dot, Table 2). Time offset is $t_0 = 2457934.9$ days.

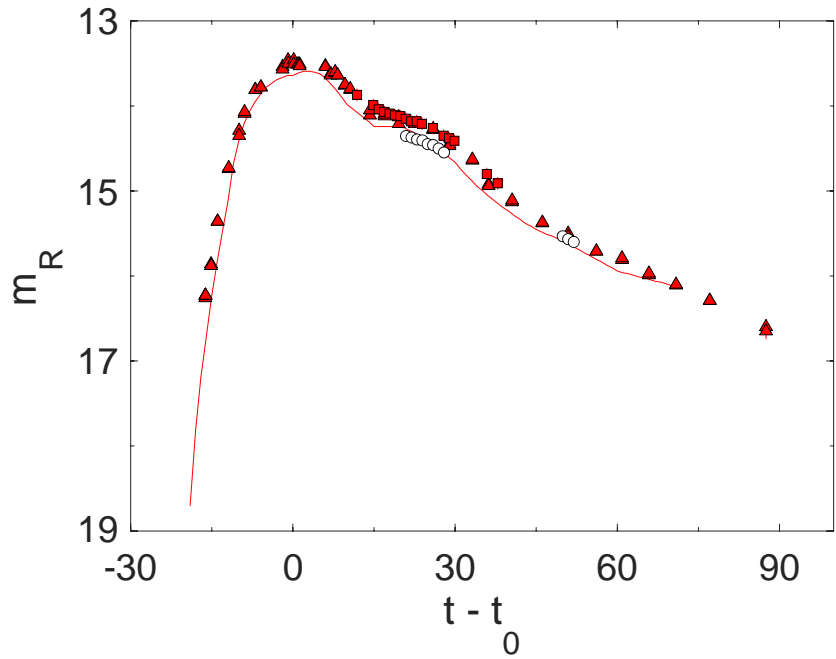

Fig. 8. SN 2017erp light curve in *R* filter. Template light curve for Type Ia SN with $z = 0$ (red line, Nugent, Kim & Perlmutter 2002) with AZT data (red squares, Brown et al. 2016), LCO data (red triangles, Brown et al. 2016) and OAUNI $m_2$ data (white dot, Table 2). Time offset is $t_0 = 2457934.9$ days.

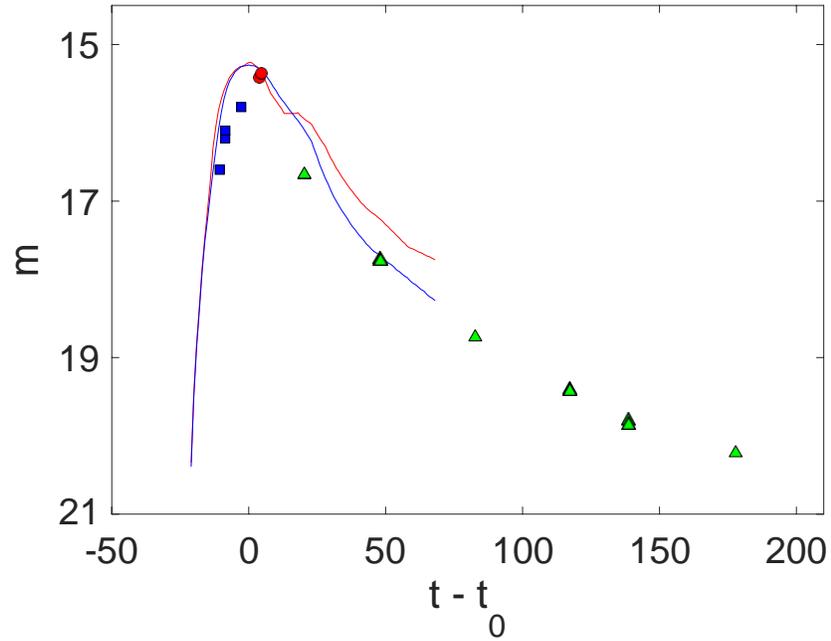

Fig. 9. AT 2017erv light curve. Template light curve for Type Ia SN in *V* (blue line) and *R* (red line) filters with *z* = 0 (Nugent, Kim & Perlmutter 2002), with *ASASSN* data (*V* filter, blue squares, Nicholls, Brimacombe & Cacella 2017), GAIA data (*G* filter, green triangles, http://gsaweb.ast.cam.ac.uk/alerts/alert/Gaia17bto/) and OAUNI data (*R* filter, red dots, Table 2). Time offset is $t_0$ = 2457926.5 days.

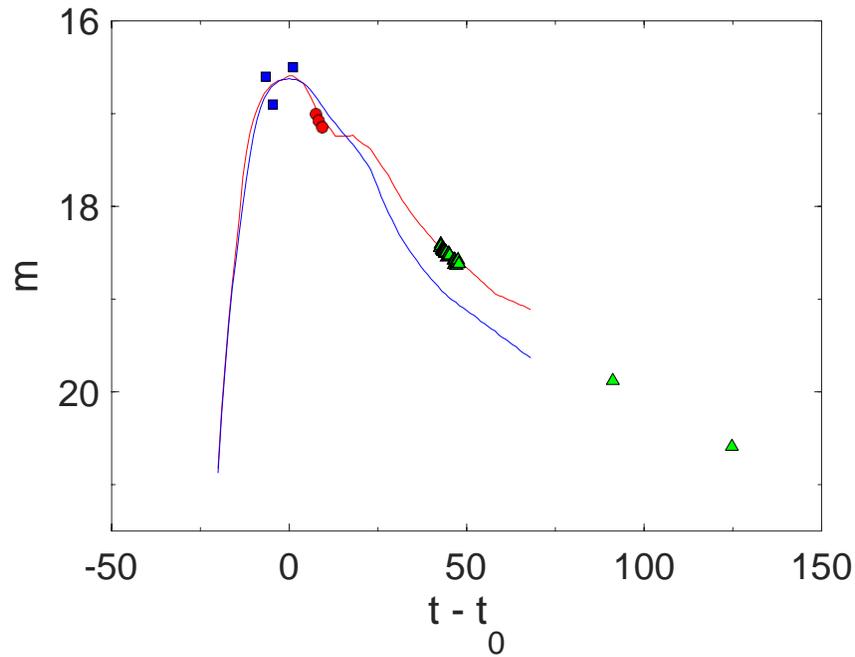

Fig. 10. AT 2017eve light curve. Template light curve for Type Ia SN in *V* (blue line) and *R* (red line) filters with $z = 0$ (Nugent, Kim & Perlmutter 2002), with *ASASSN* data (*V* filter, blue squares, Nicholls, Brimacombe & Cacella 2017), GAIA data (*G* filter, green triangles, http://gsaweb.ast.cam.ac.uk/alerts/alert/Gaia17byi/) and OAUNI data (*R* filter, red dots, Table 2). Time offset is $t_0$ = 2457922.5 days.

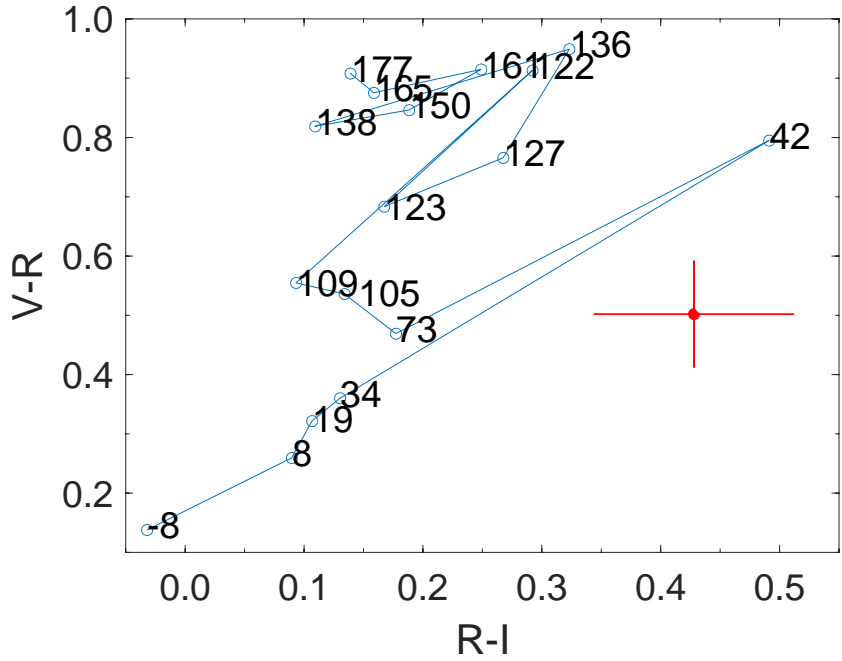

Fig. 11. $V - R$ vs. $R - I$ color diagram of a Type IIp SN with $z = 0$ (blue lines, adapted from Poznanski et al. 2002). Numbers indicate days after maximum of light. OAUNI colors for SN 2023gfo using $m_2$ calibration are also indicated (red point).

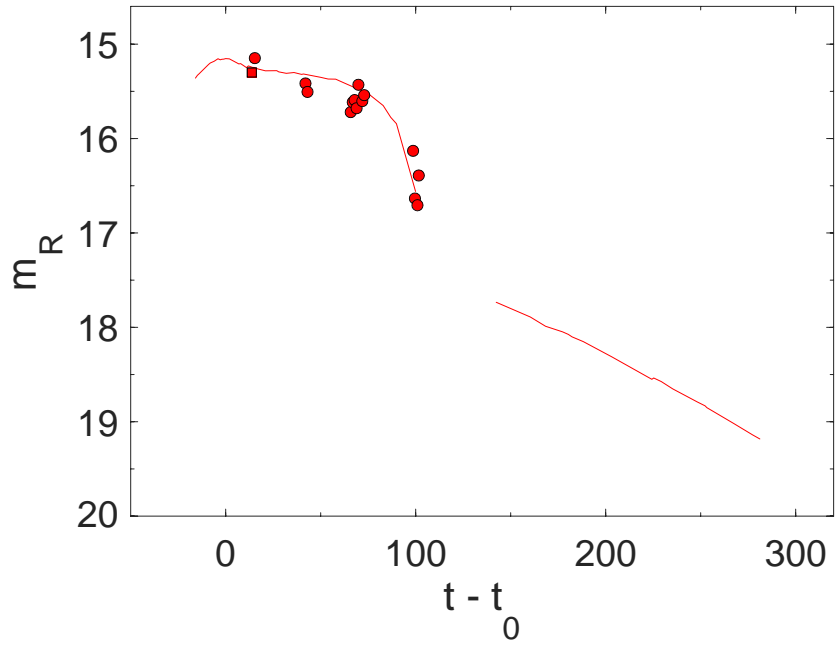

Fig. 12. SN 2023gfo light curve in the *R* filter. SN 2004et Type IIp light curve with *z* = 0.0002 (red squares, Sahu 2006), with OAUNI data (white dots, Table 2). Time offsets is $t_0$ = 2460049.7.

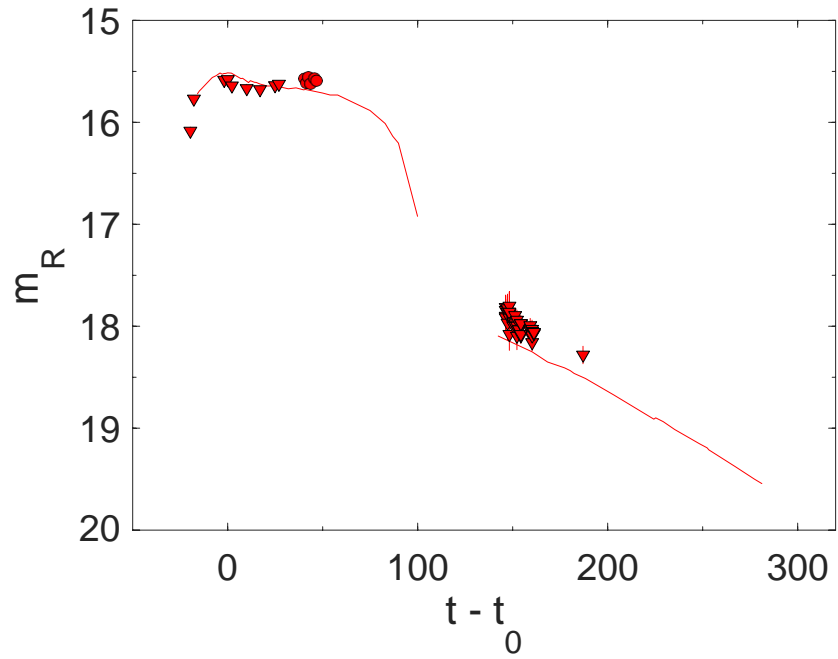

Fig. 13. SN 2023ijd light curve in the *R* filter. SN 2004et Type IIp light curve with *z* = 0.0002 (red squares, Sahu 2006), with ZTF data (red triangles, https://lasair-ztf.lsst.ac.uk/objects/ZTF23aajrmfh/), and OAUNI data (white dots, Table 2). Time offsets is $t_0$ = 2460088.9 days.

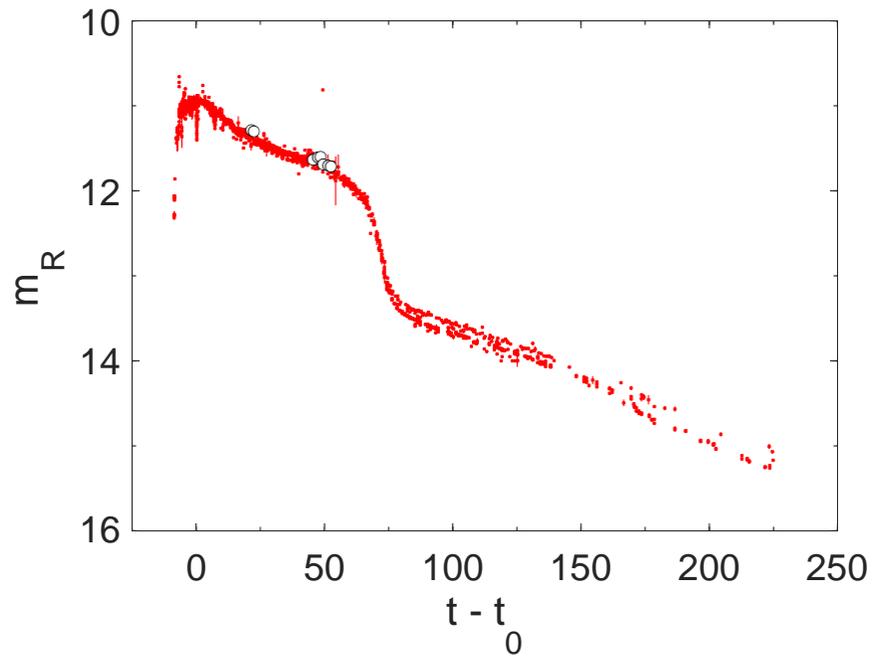

Fig. 14. SN 2023ixf light curve in *R* filter. AAVSO data (red points, www.aavso.org) and OAUNI data (white dots, Table 2). Time offset $t_0$ = 2460094.0.

## TABLE 1
### OAUNI SUPERNOVAE SAMPLE

| SN | other name | Type | RA (2000) | DEC (2000) | Discovery date (UT) | Host galaxy | $z^a$ |
|---|---|---|---|---|---|---|---|
| AT 2015dd | PSN J141[b] | Ib | 14:10:23.42 | -43:18:43.70 | 2015-12-15 | NGC 5483 | 0.005921 |
| SN 2016cvk | ASASSN-16jt | IIn-pec | 22:19:49.43 | -40:40:05.50 | 2016-06-12 | ESO 344-G21 | 0.010842 |
| AT 2016eqb | ASASSN-16hz | Ia | 23:15:45.48 | -01:20:22.73 | 2016-08-01 | 2MASX[c] | 0.02531(15) |
| SN 2017erp | | Ia | 15:09:14.90 | -11:20:03.00 | 2017-06-13 | NGC 5861 | 0.006303 |
| AT 2017erv | ASASSN-17ho | Ia | 19:18:47.10 | -84:41:50.03 | 2017-06-13 | AM 1904-844 | 0.017035 |
| AT 2017eve | ASASSN-17hq | Ia | 18:43:53.51 | -56:29:29.04 | 2017-06-19 | GALEXASC[d] | 0.031 |
| SN 2023gfo | | IIP | 13:09:39.68 | -07:50:11.75 | 2023-04-20 | NGC 4995 | 0.005834 |
| SN 2023ijd | ASASSN-23du | II | 12:36:32.47 | +11:13:19.71 | 2023-05-14 | NGC 4568 | 0.00744(10) |
| SN 2023ixf | | IIL | 14:03:38.56 | +54:18:41.94 | 2023-05-19 | M101 | 0.000811(16) |

[a] of host galaxy from SIMBAD, except for AT 2016erv (Nicholls, Brimacombe & Cacella 2017) and AT 2017eve (Uddin et al. 2017)
[b] PSN J14102342-4318437
[c] 2MASX J23154564-0120135
[d] GALEXASC J184352.21-562927.7

## TABLE 2
### OAUNI SUPERNOVAE OBSERVATION LOG

| SN (1) | date (UTC) (2) | filter (3) | N (4) | IT (s) (5) | X (6) | $Mag_1$ (7) | $Mag_2$ (8) |
|---|---|---|---|---|---|---|---|
| AT 2015dd | 2016/01/10.385 | V | 30 | 600 | 1.467 | 16.470 ± 0.060 | 16.797 ± 0.060 |
| | 2016/01/10.396 | R | 30 | 600 | 1.402 | 15.992 ± 0.033 | 15.933 ± 0.033 |
| | 2016/01/10.410 | I | 30 | 600 | 1.336 | 15.565 ± 0.037 | 15.681 ± 0.037 |
| SN 2016cvk | 2016/09/05.084 | V | 90 | 1800 | 1.375 | 15.025 ± 0.079 | 14.957 ± 0.052 |
| | 2016/09/05.099 | R | 90 | 1800 | 1.305 | 15.183 ± 0.076 | 14.911 ± 0.050 |
| | 2016/09/05.113 | I | 90 | 1800 | 1.250 | 15.247 ± 0.079 | 14.909 ± 0.065 |
| AT 2016eqb | 2016/08/09.405 | V | 45 | 900 | 1.310 | 15.806 ± 0.046 | 15.847 ± 0.046 |
| | 2016/08/09.390 | R | 45 | 900 | 1.243 | 15.806 ± 0.046 | 15.863 ± 0.046 |
| SN 2017erp | 2017/07/23.087 | V | 45 | 900 | 1.064 | 14.349 ± 0.014 | 14.424 ± 0.019 |
| | 2017/07/23.055 | R | 45 | 900 | 1.167 | 14.289 ± 0.015 | 14.355 ± 0.022 |
| | 2017/07/24.066 | V | 45 | 900 | 1.156 | 14.406 ± 0.011 | 14.442 ± 0.018 |
| | 2017/07/24.050 | R | 45 | 900 | 1.100 | 14.309 ± 0.015 | 14.372 ± 0.019 |
| | 2017/07/25.103 | V | 45 | 900 | 1.195 | 14.476 ± 0.011 | 14.502 ± 0.016 |
| | 2017/07/25.087 | R | 45 | 900 | 1.280 | 14.332 ± 0.016 | 14.395 ± 0.022 |
| | 2017/07/26.200 | V | 45 | 900 | 2.756 | 14.533 ± 0.013 | 14.544 ± 0.017 |
| | 2017/07/26.186 | R | 45 | 900 | 3.614 | 14.359 ± 0.014 | 14.405 ± 0.016 |
| | 2017/07/27.059 | V | 45 | 900 | 1.041 | 14.581 ± 0.018 | 14.611 ± 0.020 |
| | 2017/07/27.034 | R | 45 | 900 | 1.104 | 14.396 ± 0.017 | 14.450 ± 0.013 |
| | 2017/07/28.042 | V | 45 | 900 | 1.035 | 14.621 ± 0.014 | 14.656 ± 0.016 |
| | 2017/07/28.027 | R | 45 | 900 | 1.065 | 14.423 ± 0.014 | 14.461 ± 0.016 |
| | 2017/07/29.053 | V | 45 | 900 | 1.064 | 14.682 ± 0.012 | 14.706 ± 0.014 |
| | 2017/07/29.039 | R | 45 | 900 | 1.105 | 14.471 ± 0.012 | 14.501 ± 0.015 |
| | 2017/07/30.044 | V | 45 | 900 | 1.049 | 14.745 ± 0.011 | 14.759 ± 0.014 |
| | 2017/07/30.029 | R | 45 | 900 | 1.085 | 14.527 ± 0.012 | 14.548 ± 0.015 |
| | 2017/08/21.041 | V | 45 | 900 | 1.255 | 15.725 ± 0.014 | 15.695 ± 0.013 |
| | 2017/08/21.025 | R | 45 | 900 | 1.362 | 15.662 ± 0.016 | 15.532 ± 0.010 |
| | 2017/08/22.064 | V | 20 | 400 | 1.350 | 15.731 ± 0.026 | 15.733 ± 0.016 |
| | 2017/08/22.036 | R | 15 | 300 | 1.617 | 15.680 ± 0.019 | 15.568 ± 0.014 |
| | 2017/08/23.090 | V | 60 | 1200 | 1.747 | 15.780 ± 0.016 | 15.755 ± 0.018 |
| | 2017/08/23.070 | R | 60 | 1200 | 2.145 | 15.707 ± 0.026 | 15.601 ± 0.010 |
| AT 2017erv | 2017/06/26.354 | R | 45 | 900 | 3.535 | 15.423 ± 0.030 | |
| | 2017/06/27.085 | R | 45 | 900 | 3.907 | 15.372 ± 0.018 | |
| AT 2017eve | 2017/06/26.386 | R | 45 | 900 | 2.138 | 17.010 ± 0.032 | |
| | 2017/06/27.156 | R | 45 | 900 | 1.503 | 17.077 ± 0.033 | |
| | 2017/06/28.158 | R | 90 | 1800 | 1.485 | 17.145 ± 0.028 | |
| SN 2023gfo | 2023/05/24.228 | V | 45 | 900 | 1.637 | 15.716 ± 0.025 | 15.651 ± 0.048 |
| | 2023/05/24.242 | R | 45 | 900 | 1.866 | 15.149 ± 0.022 | 15.149 ± 0.042 |
| | 2023/05/24.212 | I | 45 | 900 | 1.457 | 14.698 ± 0.021 | 14.721 ± 0.042 |
| | 2023/06/20.079 | R | 60 | 1200 | 1.108 | 15.417 ± 0.025 | |
| | 2023/06/21.091 | R | 80 | 1600 | 1.162 | 15.507 ± 0.025 | |
| | 2023/07/14.022 | R | 70 | 1400 | 1.144 | 15.720 ± 0.025 | |
| | 2023/07/15.045 | R | 70 | 1400 | 1.275 | 15.616 ± 0.025 | |
| | 2023/07/16.084 | R | 70 | 1400 | 1.663 | 15.592 ± 0.023 | |
| | 2023/07/17.079 | R | 70 | 1400 | 1.633 | 15.680 ± 0.025 | |
| | 2023/07/18.105 | R | 70 | 1400 | 2.153 | 15.431 ± 0.023 | |
| | 2023/07/20.102 | R | 70 | 1400 | 2.231 | 15.604 ± 0.026 | |
| | 2023/07/21.107 | R | 70 | 1400 | 2.465 | 15.540 ± 0.024 | |
| | 2023/08/16.057 | R | 70 | 1400 | 3.761 | 16.131 ± 0.032 | |
| | 2023/08/17.039 | R | 90 | 1400 | 2.801 | 16.636 ± 0.038 | |
| | 2023/08/18.038 | R | 90 | 1400 | 2.903 | 16.706 ± 0.038 | |
| | 2023/08/19.038 | R | 90 | 1400 | 2.768 | 16.391 ± 0.034 | |
| SN 2023ijd | 2023/07/15.070 | R | 70 | 1400 | 2.554 | 15.571 ± 0.022 | |
| | 2023/07/16.054 | R | 70 | 1400 | 2.174 | 15.613 ± 0.018 | |
| | 2023/07/17.050 | R | 70 | 1400 | 2.135 | 15.557 ± 0.018 | |
| | 2023/07/18.071 | R | 70 | 1400 | 2.908 | 15.621 ± 0.022 | |
| | 2023/07/20.070 | R | 70 | 1400 | 3.200 | 15.568 ± 0.013 | |
| | 2023/07/21.079 | R | 70 | 1400 | 4.045 | 15.591 ± 0.018 | |
| SN 2023ixf | 2023/06/20.021 | R | 54 | 1080 | 2.499 | 11.285 ± 0.025 | |
| | 2023/06/21.022 | R | 50 | 1000 | 2.483 | 11.301 ± 0.023 | |
| | 2023/07/14.050 | R | 53 | 1060 | 2.804 | 11.635 ± 0.024 | |
| | 2023/07/16.025 | R | 50 | 1000 | 2.615 | 11.606 ± 0.024 | |
| | 2023/07/17.018 | R | 50 | 1000 | 2.585 | 11.600 ± 0.029 | |
| | 2023/07/18.031 | R | 50 | 1000 | 2.715 | 11.690 ± 0.023 | |
| | 2023/07/20.038 | R | 53 | 1060 | 2.856 | 11.704 ± 0.026 | |
| | 2023/07/21.052 | R | 50 | 1000 | 3.131 | 11.714 ± 0.027 | |

TABLE 3

UCAC4 COMPARISON STARS

| UCAC4 name | $V_{cat}$ (mag) (1) | $R_{cat}$ (mag) (2) | $I_{cat}$ (mag) (3) |
|---|---|---|---|
| \multicolumn{4}{c}{AT 2015dd} | | | |
| 235-072564 | 11.356 | 10.968 | 10.607 |
| 235-072546 | 12.711 | 12.394 | 12.108 |
| 234-070894 | 9.800 | 9.879 | 9.879 |
| 234-070931 | 10.706 | 10.353 | 9.965 |
| 234-070956 | 12.923 | 12.591 | 12.292 |
| \multicolumn{4}{c}{SN 2016cvk} | | | |
| 247-183586 | 12.291 | 12.176 | 12.060 |
| 247-183587 | 15.491 | 15.333 | 15.205 |
| 247-183594 | 14.920 | 14.771 | 14.651 |
| 247-183606 | 14.646 | 14.477 | 14.352 |
| 247-183607 | 13.416 | 13.317 | 13.222 |
| 247-183608 | 14.249 | 14.099 | 13.966 |
| 248-191420 | 12.131 | 11.962 | 11.834 |
| 248-191421 | 15.182 | 15.036 | 14.954 |
| \multicolumn{4}{c}{AT 2016eqb} | | | |
| 444-131450 | 16.501 | 15.924 | - |
| 444-131442 | 15.745 | 15.506 | - |
| 445-136436 | 16.153 | 16.061 | - |
| 445-136441 | 16.506 | 16.386 | - |
| 445-136458 | 15.839 | 15.539 | - |
| 445-136465 | 16.083 | 15.858 | - |
| \multicolumn{4}{c}{SN 2017erp} | | | |
| 394-058196 | 14.691 | 14.464 | - |
| 394-058202 | 14.800 | 14.535 | - |
| 394-058177 | 14.867 | 14.717 | - |
| 393-061368 | 15.085 | 14.916 | - |
| 394-058204 | 15.196 | 14.986 | - |
| 394-058168 | 15.204 | 14.870 | - |
| \multicolumn{4}{c}{AT 2017erv} | | | |
| 027-009815 | - | 12.024 | - |
| 027-009769 | - | 12.309 | - |
| 027-009780 | - | 12.704 | - |
| 027-009785 | - | 12.884 | - |
| 027-009751 | - | 12.937 | - |
| 027-009784 | - | 12.985 | - |
| 027-009787 | - | 13.197 | - |
| 027-009792 | - | 13.255 | - |
| 027-009754 | - | 13.413 | - |
| 027-009747 | - | 13.431 | - |
| 027-009800 | - | 13.443 | - |
| 027-009768 | - | 13.487 | - |
| 027-009794 | - | 13.543 | - |
| 027-009807 | - | 13.547 | - |
| 027-009809 | - | 13.754 | - |
| 027-009755 | - | 13.775 | - |
| \multicolumn{4}{c}{AT 2017eve} | | | |
| 168-205741 | - | 12.147 | - |
| 168-205790 | - | 12.191 | - |
| 168-205740 | - | 12.395 | - |
| 168-205715 | - | 12.493 | - |
| 168-205795 | - | 12.778 | - |
| 168-205776 | - | 13.051 | - |
| 168-205727 | - | 13.226 | - |
| 168-205789 | - | 13.628 | - |
| 168-205806 | - | 13.716 | - |
| 169-196343 | - | 13.748 | - |
| 168-205754 | - | 13.781 | - |
| 169-196291 | - | 13.947 | - |
| 168-205770 | - | 13.952 | - |
| 168-205798 | - | 13.996 | - |
| 169-196333 | - | 14.000 | - |
| \multicolumn{4}{c}{SN 2023gfo} | | | |
| 412-054666 | 14.292 | 14.114 | 14.018 |
| 412-054667 | 14.279 | 14.121 | 14.030 |
| 412-054681 | 15.068 | 14.674 | 14.433 |
| 411-055301 | 15.112 | 14.930 | 14.793 |
| 411-055298 | 14.656 | 14.203 | 13.907 |
| 411-055299 | 15.053 | 14.781 | 14.604 |
| \multicolumn{4}{c}{SN 2023ijd} | | | |
| 506-053220 | - | 15.925 | - |
| 506-053224 | - | 15.901 | - |
| 506-054413 | - | 16.161 | - |
| 506-054416 | - | 15.760 | - |
| 506-054417 | - | 15.790 | - |
| 506-054418 | - | 16.008 | - |
| \multicolumn{4}{c}{SN 2023ixf} | | | |
| 723-053563 | - | 15.400 | - |
| 723-053565 | - | 14.719 | - |
| 723-053569 | - | 14.582 | - |
| 722-053112 | - | 15.410 | - |
| 722-053102 | - | 14.245 | - |
| 722-053103 | - | 15.298 | - |

TABLE 4

AVAILABLE LITERATURE LIGHT CURVE DATA

| SN | date (UT) | G | V | R | reference |
|---|---|---|---|---|---|
| SN 2016cvk | 2016-08-31.09 | - | 16.2 | - | (a) |
|  | 2016-09-10.60 | - | 14.69 | - | (b) |
|  | 2016-10-22.42 | - | 16.04 | - | (b) |
| AT 2016eqb | 2016-08-01.35 | - | 16.8 | - | (a) |
|  | 2016-08-03.27 | - | 16.2 | - | (a) |
|  | 2016-08-05.31 | - | 16.5 | - | (a) |
|  | 2016-08-06.18 | - | 16.5 | - | (a) |
| AT 2017erv | 2017-06-11.23 | - | 16.6 | - | (c) |
|  | 2017-06-13.31 | - | 16.2 | - | (c) |
|  | 2017-06-13.38 | - | 16.1 | - | (c) |
|  | 2017-06-19.25 | - | 15.8 | - | (c) |
|  | 2017-07-13.04 | 16.67 | - | - | (d) |
|  | 2017-07-13.08 | 16.66 | - | - | (d) |
|  | 2017-08-10.08 | 17.75 | - | - | (d) |
|  | 2017-08-10.29 | 17.77 | - | - | (d) |
|  | 2017-09-14.33 | 18.74 | - | - | (d) |
|  | 2017-10-19.25 | 19.41 | - | - | (d) |
|  | 2017-10-19.29 | 19.43 | - | - | (d) |
|  | 2017-11-09.96 | 19.81 | - | - | (d) |
|  | 2017-11-10.04 | 19.86 | - | - | (d) |
|  | 2017-12-19.67 | 20.22 | - | - | (d) |
| AT 2017eve | 2017-06-11.18 | - | 16.6 | - | (c) |
|  | 2017-06-13.26 | - | 16.9 | - | (c) |
|  | 2017-06-19.14 | - | 16.5 | - | (c) |
|  | 2017-08-01.29 | 18.45 | - | - | (e) |
|  | 2017-08-01.54 | 18.41 | - | - | (e) |
|  | 2017-08-01.63 | 18.43 | - | - | (e) |
|  | 2017-08-01.88 | 18.47 | - | - | (e) |
|  | 2017-08-01.88 | 18.47 | - | - | (e) |
|  | 2017-08-02.13 | 18.48 | - | - | (e) |
|  | 2017-08-02.29 | 18.49 | - | - | (e) |
|  | 2017-08-02.38 | 18.47 | - | - | (e) |
|  | 2017-08-02.54 | 18.48 | - | - | (e) |
|  | 2017-08-02.63 | 18.48 | - | - | (e) |
|  | 2017-08-02.79 | 18.50 | - | - | (e) |
|  | 2017-08-02.88 | 18.51 | - | - | (e) |
|  | 2017-08-03.04 | 18.50 | - | - | (e) |
|  | 2017-08-03.13 | 18.50 | - | - | (e) |
|  | 2017-08-03.29 | 18.55 | - | - | (e) |
|  | 2017-08-03.63 | 18.54 | - | - | (e) |
|  | 2017-08-03.88 | 18.51 | - | - | (e) |
|  | 2017-08-04.13 | 18.53 | - | - | (e) |
|  | 2017-08-05.29 | 18.58 | - | - | (e) |
|  | 2017-08-05.38 | 18.63 | - | - | (e) |
|  | 2017-08-05.54 | 18.58 | - | - | (e) |
|  | 2017-08-05.63 | 18.58 | - | - | (e) |
|  | 2017-08-05.79 | 18.61 | - | - | (e) |
|  | 2017-08-05.88 | 18.60 | - | - | (e) |
|  | 2017-08-06.04 | 18.64 | - | - | (e) |
|  | 2017-08-06.29 | 18.62 | - | - | (e) |
|  | 2017-08-06.38 | 18.61 | - | - | (e) |
|  | 2017-08-06.54 | 18.58 | - | - | (e) |
|  | 2017-08-06.79 | 18.62 | - | - | (e) |
|  | 2017-09-20.54 | 19.88 | - | - | (e) |
|  | 2017-10-25.08 | 20.59 | - | - | (e) |
| SN 2023ijd | 2023-07-01.21 | - | - | 15.649±0.031 | (f) |
|  | 2023-06-29.17 | - | - | 15.661±0.030 | (f) |
|  | 2023-06-21.21 | - | - | 15.701±0.034 | (f) |
|  | 2023-06-14.17 | - | - | 15.689±0.041 | (f) |
|  | 2023-06-06.29 | - | - | 15.667±0.026 | (f) |
|  | 2023-06-04.21 | - | - | 15.601±0.029 | (f) |
|  | 2023-06-02.21 | - | - | 15.606±0.031 | (f) |
|  | 2023-05-17.21 | - | - | 15.796±0.033 | (f) |
|  | 2023-05-15.29 | - | - | 16.121±0.039 | (f) |

[a] Brimacombe et al. (2016)
[b] http://ooruri.kusastro.kyoto-u.ac.jp/mailarchive/vsnet-recent-sn/6612
[c] Nicholls, Brimacombe & Cacella (2017)
[d] http://gsaweb.ast.cam.ac.uk/alerts/alert/Gaia17bto/
[e] http://gsaweb.ast.cam.ac.uk/alerts/alert/Gaia17byi/
[f] https://lasair-ztf.lsst.ac.uk/objects/ZTF23aajrmfh/

TABLE 5

OAUNI SUPERNOVAE PARAMETERS

| SN | filter | Peak date (JD) | $m$ (mag) | residual (mag) ($\times 10^{-4}$) |
|---|---|---|---|---|
| AT 2015dd | - | 2457374.5 | - | - |
| SN 2016cvk | R | 2457648.5 | 14.7197 | 1.0569 |
| AT 2016eqb | V | 2457607.5 | 16.2130 | 2.1325 |
| SN 2017erp | R | 2457934.9 | 13.5922 | 3.0811 |
|  | V | 2457934.9 | 13.4813 | 2.6995 |
| AT 2017erv | R | 2457926.5 | 15.2615 | 2.8972 |
| AT 2017eve | R | 2457922.5 | 16.6213 | 3.5673 |
| SN 2023gfo | R | 2460049.7 | 15.1622* | 2.8806 |
| SN 2023ijd | R | 2460088.9 | 15.6157* | 2.7022 |
| SN 2023ixf | R | 2460094.0 | - | - |

*SN 2004et was used as a template